\documentclass[twocolumn,letterpaper]{IEEEAerospaceCLS}  % only supports two-column, letterpaper format

\usepackage[T1]{fontenc}
\usepackage[active]{srcltx} % SRC Specials for DVI Searching
\usepackage{amsmath}
\usepackage{color}
\usepackage{amssymb}
\usepackage{color}
\usepackage[final]{graphicx}
\usepackage{url}
\usepackage{enumerate}
\usepackage{epsfig}
\usepackage{latexsym}
\usepackage{euscript}
\usepackage{mathrsfs}
\usepackage{psfrag}
\usepackage{cite}
\usepackage{url}
\usepackage[english]{babel}
\usepackage[linesnumbered,ruled,vlined]{algorithm2e}
\usepackage{fancyvrb}
\usepackage{fvextra}
\usepackage{booktabs}
\usepackage{tikz}
\usepackage{textcomp}
\usepackage{lipsum}
\usepackage[T1]{fontenc}
\usepackage{booktabs}
\usepackage{multirow}
\usepackage{siunitx}

\newcommand{\Es}{{\mathbb{E}}}
\newcommand{\ignore}[1]{}  % {} empty inside = %% comment
\newcommand{\Rset}{\mathbb{R}}
\newcommand{\Zset}{\mathbb{Z}}
\newcommand{\Nset}{\mathbb{N}}
\newcommand{\eqdef}{\triangleq}

\newcommand{\code}[1]{\texttt{#1}}
\newcommand{\Ts}{T_{\text{s}}}
\newcommand{\fs}{f_{\text{s}}}
\newcommand{\Mneg}{M^{-}}
\newcommand{\Mpos}{M^{+}}

\def\be#1\ee{\begin{equation}#1\end{equation}}
\def\barr#1\earr{\begin{align}#1\end{align}}

\newcommand\acceptedtext{%
	\footnotesize This article has been accepted for publication in the proceedings of this conference, but has not been fully edited.\\ Content may change prior to final publication. \\
	Citation information: DOI 10.1109/AERO58975.2024.10521369, 2024 IEEE Aerospace Conference.}
\newcommand\acceptednotice{%
	\begin{tikzpicture}[remember picture,overlay]
		\node[anchor=north,yshift=-6pt] at (current page.north) {%
			\begin{minipage}{\textwidth}
				\center \acceptedtext
		\end{minipage}};
	\end{tikzpicture}%
}

\begin{document}

\title{\code{CycloDSP}: A cyclostationary signal analysis \\ tool for GNU Radio}

\author{%
Donatella Darsena\\ 
Dept. of Electrical Engineering\\ 
and Information Technology \\ 
University Federico II\\ 
Naples I-80125\\
darsena@unina.it
\and
Giacinto Gelli\\ 
Dept. of Electrical Engineering\\ 
and Information Technology \\ 
University Federico II\\ 
Naples I-80125\\
gelli@unina.it
\and 
Ivan Iudice\\
Dept. of Reliability \& Security \\
Italian Aerospace Research Centre \\
(CIRA) \\
Capua I-81043 \\
i.iudice@cira.it
%%%% IMPORTANT: Use the correct copyright information--IEEE, Crown, or U.S. government. %%%%%
\thanks{\footnotesize 979-8-3503-0462-6/24/$\$31.00$ \copyright2024 IEEE.
Personal use is permitted, but republication/redistribution requires IEEE permission.
See https://www.ieee.org/publications/rights/index.html for more information.}              % This creates the copyright info that is the correct 2024 data.
%\thanks{{U.S. Government work not protected by U.S. copyright}}         % Use this copyright notice only if you are employed by the U.S. Government.
%\thanks{{979-8-3503-0462-6/24/$\$31.00$ \copyright2024 Crown}}          % Use this copyright notice only if you are employed by a crown government (e.g., Canada, UK, Australia).
%\thanks{{979-8-3503-0462-6/24/$\$31.00$ \copyright2024 European Union}}    % Use this copyright notice is you are employed by the European Union.
}

\maketitle
\acceptednotice

\thispagestyle{plain}
\pagestyle{plain}
 
\maketitle

\thispagestyle{plain}
\pagestyle{plain}

\begin{abstract}
In this paper, we present a first attempt 
to incorporate in the GNU Radio ecosystem
a tool called \code{CycloDSP}, 
devoted to the analysis of 
complex-valued cyclostationary signals.
Such signals are ubiquitous in communication and signal processing,
exhibiting periodic or almost periodic statistics that are 
characterized by a countable set of {cycle frequencies}, 
which are related to the main signal periodicities.
Common cycle frequencies
for modulated signals are multiple 
of the baud-rate and/or combination
of the carrier frequency and baud-rate. 
%
%Second-order CS representative 
%functions in the time and frequency domain 
%are the cyclic correlation and the spectral correlation function, respectively.
%
Conventional estimation strategies for cyclostationary 
signal analysis typically 
exhibit a high computational burden.
Many approaches aimed at reducing complexity
exploit fast Fourier transform (FFT) algorithms,
which are very efficient for batch data but
are not suited for continuously streaming data, 
typically encountered in software-defined radio (SDR)
applications.
%Moreover, common spectral
%analysis tools for GNU Radio, such as \code{gr-specest},  
%are not able to evaluate the spectral correlation function. 
%
The aim of this paper is to develop 
an out-of-tree (OOT) GNU Radio module 
containing a set of building block functions,
aimed at estimating functions, such as
the cyclic correlation functions,
typically employed for cyclostationary signal analysis.
The proposed implementation must be 
designed so as to ensure high efficiency 
in processing continuous data 
streams of complex samples,
both in terms of computational load and data storage, 
in order to be executed on general purpose computers.
We tested our implementation by estimating
in real-time the second-order cyclic 
statistics of a Gaussian minimum-shift keying (GMSK)
modulated signal, which belongs to a class
of signals commonly employed in aeronautical 
telemetry datalinks.
\end{abstract} 

\tableofcontents

%======================================================================
\section{Introduction}
\label{sec:intro}

According to Mitola \cite{Mitola1992}, the term
\emph{software-defined radio} (SDR) 
identifies a class of radios that
can be reprogrammed 
and reconfigured through
software.
The adoption of SDR low-cost 
development platforms is envisioned as a key enabler for
many aerospace applications, such as unmanned aerial vehicles (UAVs) \cite{Yue2018},
small satellites \cite{Ceylan2016}, or deep space communications \cite{Pugh2017}.
Indeed, in operating scenarios  
characterized by a lifetime of fifteen-year or more, 
the SDR approach emerges as the preferred solution to 
overcome the obsolescence problem, 
since it allows one to design flexible and reconfigurable 
payloads, without a significant increase in 
complexity and cost \cite{Angeletti2014}.

\emph{Cyclostationary signals} are ubiquitous 
in communications and signal processing \cite{Gardner1975}.
They exhibit \textit{periodic} or \textit{almost periodic} \cite{Corduneanu1968} 
statistics, which derive
from periodic signal processing  
operations carried  out at the transmitter, such as
sampling, scanning, modulation, multiplexing, interleaving, coding, 
etc.
As a consequence,  
the statistical functions characterizing the signal, 
such as, e,g., its autocorrelation function (ACF), 
can be expanded in a generalized Fourier series, 
whose coefficients, called \textit{cyclic statistics} 
(e.g., \textit{cyclic ACFs} in the particular case)
contain useful information  about the underlying 
periodicities characterizing 
the signal.
Indeed, many algorithms have been developed,
which employ cyclic statistics 
for many scopes, including time-varying filtering 
\cite{Gardner1993}, detection/interception \cite{Gardner1993a}, 
array processing \cite{Agee1990}, synchronization \cite{Gini1998}, source location \cite{Gelli1996}, 
modulation recognition \cite{Spooner2017}, and many others.
One of the advantages of cyclic-based approaches 
is their \textit{signal selectivity} \cite{Schell1989},
i.e., their insensitivity to 
disturbance and/or noise contributions.

Among SDR software platforms, 
\textit{GNU Radio} \cite{GNU-Radio} is an open-source framework,
widely adopted by industrial and academic researchers, 
as well as amateurs 
and practitioners, which
allows for fast prototyping of communication and 
signal processing solutions.
SDR solutions based on GNU Radio 
for \emph{real-time} 
analysis of cyclostationary signals  
have been presented
in some papers, mainly oriented 
to \textit{spectrum sensing}
for cognitive radio applications
(see \cite{Manco2022}  for a recent survey on this topic).
In \cite{Zhou2012} a GNU Radio solution 
for real-time estimation of cyclic features 
in the frequency domain is proposed, 
with particular attention to reducing
the computational burden 
by a two-stage approach, 
composed by a coarse 
estimation stage,
followed by a fine estimation stage carried out only
in close proximity to the cycle features 
determined in the first stage.
The main task of the algorithm proposed in \cite{Zhou2012} 
is to extract from the estimated cyclostationary features
some signal parameters of interest,
such as, e.g., the carrier frequency or the symbol rate.

In \cite{Baldini2012,Paisana2012},
spectrum sensing algorithms 
based on cyclostationary 
features have been implemented
and tested using GNU Radio and the Universal Software Radio 
Peripheral (USRP) platform.
In \cite{Aziz2014} 
the complexity and performance
of an algorithm exploiting the sparsity
of the cyclic ACF is analyzed
using GNU Radio/USRP over a real radio channel.
In \cite{Nafkha2014,Nafkha2015,Nafkha2015a} 
several spectrum sensing 
algorithms, including those based on cyclostationary 
features, have been implemented and tested using GNU Radio/USRP.
In \cite{Laghate2017} cyclostationary 
features are used to
perform hierarchical modulation classification, and the resulting algorithms are
implemented in C++ and tested using 
GNU Radio/USRP.
Recently, in \cite{Abed2021} cyclostationary
feature detectors for spectrum sensing are implemented  
in the GNU Radio/USRP environment, 
which are characterized by 
a reduced design complexity in comparison 
to other solutions in the literature.

The previous GNU Radio implementations are all devoted to specific 
applications, such as
spectrum sensing or modulation classification.
{As a consequence, 
they employ specific time-domain or frequency-domain
cyclic estimators, optimized
to suit the needs of the 
applications at hand.}
To the best of our knowledge, however,
there is no general-purpose tool 
for real-time analysis of  complex-valued cyclostationary signals 
implemented in the GNU Radio framework, 
{considering
both time-domain and frequency-domain features,} 
as well as conventional and \textit{conjugate} cyclic features
(see Section~\ref{sec:ACS-signals}).
Thus, in this paper, we consider the development
of an out-of-tree (OOT) GNU Radio module, called
\code{CycloDSP}, containing a set of building block functions
aimed at real-time estimation and visualization
of the cyclic features of radio signals.
The proposed implementation is specifically designed
so as to work on continuous data streams,
in order to be executed in real-time environments.
Our work is a first step towards development 
of an extensible  library of highly optimized algorithms 
for cyclostationary signal analysis,
both in time and frequency domains, 
which can be used as a tool 
to implement more sophisticated
real-time algorithms in aerospace applications
and other fields.
We tested our module using GNU Radio/USRP, 
by estimating in real-time the cyclic statistics
of a \textit{Gaussian minimum-shift keying}
(GMSK) modulated signal, which is a particular form
of \emph{continuous-phase modulated} (CPM) signals,
widely utilized in aeronautical and space datalinks.

The paper is organized as follows.
In Section \ref{sec:ACS-signals} we recall the main definitions 
and functions characterizing second-order wide-sense cyclostationary signals.
In Section \ref{sec:GMSK} a brief description of
CPM and GMSK modulation formats is provided, 
together with a discussion on their cyclic features.
Section \ref{sec:GR-implementation} contains
a high-level description of the GNU Radio implementation
of the \code{CycloDSP} OOT module.
In Section \ref{sec:testbed} an experimental testbed
for validating the proposed implementation is described,
using GNU Radio/USRP for real-time generation and estimation
of the cyclic features of a GMSK radio signal.
Conclusions are drawn in Section \ref{sec:concl}, 
together with some discussion for future work.

%===============================================================
\section{Wide-sense cyclostationary signals}
\label{sec:ACS-signals}

Let us consider a  complex-valued%
\footnote{Complex-valued signals arise naturally in SDR applications,
which usually work with in-phase (I) and quadrature-phase
(Q) components, so called \textit{complex envelope}.}
random  signal $x(t)$, whose second-order moments are given by
\begin{align}
R_{xx}(t,\tau) & \triangleq \Es[x(t) \, x^{*}(t-\tau)]  \label{eq:ACF-def} \\
R_{xx{*}}(t,\tau) & \triangleq \Es[x(t) \, x(t-\tau)]  \label{eq:CCF-def} 
\end{align}
with $\Es[\cdot]$ denoting the statistical 
average operator and $*$ denoting 
the complex conjugation.
In particular, eq.~\eqref{eq:ACF-def} represents
the \emph{autocorrelation function} (ACF),
whereas \eqref{eq:CCF-def} is referred to as the
\emph{conjugate correlation function} (CCF).

A finite-power signal $x(t)$ is said 
to be second-order \emph{almost-cyclostationary} 
(ACS) in the wide-sense \cite{Napolitano2012} 
if its ACF, being \textit{almost periodic}  
\cite{Corduneanu1968} in $t$,  
can be expanded (with respect to $t$)
in a generalized Fourier series:
\be
R_{xx}(t,\tau) = \sum_{\alpha \in \mathcal{A}} R^{\alpha}_{xx}(\tau) \, e^{j 2 \pi \alpha t}
\label{eq:ACF-theo}
\ee
where $\mathcal{A} \subset\Rset$ is the countable set%
\footnote{
$\Rset$ and $\Zset$ are the fields of
real and integer numbers, respectively.}
of (possibly incommensurate) \emph{cycle frequencies} $\alpha$, 
whereas each Fourier coefficient
\be
R^{\alpha}_{xx}(\tau) \triangleq 
\langle R_{xx}(t,\tau) \, e^{-j 2 \pi \alpha t} \rangle 
\label{eq:cycACF-theo}
\ee
is referred to as the \emph{cyclic} ACF 
at cycle frequency $\alpha$,
with 
\be
\langle x(t) \rangle 
\triangleq \lim_{Z \rightarrow +\infty} \frac{1}{2Z} \, \int_{-Z}^{Z} 
x(t) \text{d}t
\ee
denoting the infinite-time average operator.
In the special case where the ACF $R_{xx}(t,\tau)$ 
is periodic in $t$ with period $T_0$, i.e.,
the conventional Fourier series expansion holds:
\be
R_{xx}(t,\tau) = \sum_{k=-\infty}^{+\infty} R^{k/{T_0}}_{xx}(\tau) \, e^{j 2 \pi \frac{k}{T_0} t}
\ee
the process $x(t)$ is said to be second-order wide-sense \emph{cyclostationary} 
with period $T_0$.

Similarly, a finite-power signal $x(t)$ is said to be
second-order almost \emph{conjugate cyclostationary} in the wide-sense
if its CCF \eqref{eq:CCF-def}  
can be expanded (with respect to $t$) 
in a generalized Fourier series
as in \eqref{eq:ACF-theo}, 
where $R^{\beta}_{xx^*}(\tau)$
\textit{in lieu}
of $R^{\alpha}_{xx}(\tau)$
denotes the \emph{cyclic} CCF 
at \textit{conjugate} cycle frequency 
$\beta \in \mathcal{B} \subset \Rset$, whose expression
can be obtained by \eqref{eq:cycACF-theo}
by replacing
$\alpha$ with $\beta$ and  $x^*(t-\tau)$ with $x(t-\tau)$, respectively.

Also in this case, if all conjugate cycle 
frequencies $\beta$ can be expressed as a multiple of some 
$1/{\overline{T}_0}$, that is 
${\mathcal B} \equiv \{ m/\overline{T}_0 \} $ for $m \in \mathbb{Z}$, 
the process $x(t)$ is said to be second-order wide-sense 
\textit{conjugate cyclostationary} with period $\overline{T}_0$.

A natural estimator of
the cyclic ACF \eqref{eq:cycACF-theo}, 
based on the observation of $x(t)$ 
for $t \in [t_0-Z/2,t_0+Z/2]$, 
is given \cite{Napolitano2012} by
%the finite time-average
%of $x(t)\, x^*(t-\tau) \, e^{-j 2 \pi \alpha t}$:
%
\begin{multline}
\widehat{R}_{xx}^\alpha(\tau)=
\frac{1}{Z} 
\int_{t_0-\frac{Z}{2}}^{t_0+\frac{Z}{2}} x(t) \, x^*(t-\tau) \\
\times \text{rect}\left( \frac{t-\tau-t_0}{Z}\right) \, e^{-j 2 \pi \alpha t} \text{d}t 
\label{eq:est-ACF}
\end{multline}
where the rectangular window%
\footnote{The elementary rectangular window
is defined as $\text{rect}(t)=1$ for $|t| \le 1/2$, and zero elsewhere.}
takes into account the fact that
$x(t-\tau)$ has been observed only for $(t-\tau) \in [t_0-Z/2,t_0+Z/2]$.
The effect of such rectangular windowing becomes negligible 
as the data-record length $Z$ is much larger than 
$|\tau|$.

%Along the same line of \eqref{eq:est-ACF}, 
Likewise, an estimator of the cyclic CCF 
$R^{\beta}_{xx^*}(\tau)$
can be obtained from \eqref{eq:est-ACF}
by replacing $\alpha$ with $\beta$ and 
$x^*(t-\tau)$ with $x(t-\tau)$, respectively.
%
%is given by
%\begin{multline}
%\widehat{R}_{xx^*}^\beta(\tau)=
%\frac{1}{Z} 
%\int_{t_0-\frac{Z}{2}}^{t_0+\frac{Z}{2}} x(t) \, x(t-\tau) \\
%\times \text{rect}\left( \frac{t-\tau-t_0}{Z}\right)\, e^{-j 2 \pi \beta t} \text{d}t \: .
%\label{eq:est-CCF}
%\end{multline}
%
It can be proven 
that, under mild conditions (see e.g. \cite{Napolitano2012}), 
such cyclic ACF and CCF estimators  
are asymptotically (for $Z\to+\infty$) unbiased and 
mean-square consistent estimators 
of the cyclic ACF and CCF, respectively.

Finally, considering the discrete-time signal 
$x[n] \eqdef x(n \Ts)$
obtained by sampling, 
with sampling frequency $\fs = 1/\Ts$,
its continuous-time counterpart $x(t)$, 
the two estimators of the cyclic ACF 
and CCF become, respectively,
\begin{align}
    \widehat{R}_{xx}^{\tilde{\alpha}}[m] &= \frac{1}{2 N+1} \sum_{n=-N}^{N} x[n] x^*[n-m] \, e^{-j 2 \pi \tilde{\alpha} n} \\
    \widehat{R}_{xx^*}^{\tilde{\beta}}[m] &= \frac{1}{2 N+1} \sum_{n=-N}^{N} x[n] x[n-m] \, e^{-j 2 \pi \tilde{\beta} n}
\end{align}
where $\tilde{\alpha} \eqdef \alpha / {\fs}$ and $\tilde{\beta} \eqdef \beta / {\fs}$ 
denote the \textit{normalized}
cycle and conjugate cycle frequencies, 
respectively, 
and the sample-size $2N+1$ 
is assumed to be much larger
than the discrete-time lag $|m|$ in order 
to neglect the effect of 
rectangular windowing.

%---------------------------------------------------------------
\section{CPM and GMSK modulation formats}
\label{sec:GMSK}

The GMSK modulation format
belongs to the family of CPM 
\cite{Anderson2013} signals.
Due to their many favorable properties, 
such as robustness to nonlinear amplification,
high spectral efficiency, and noise robustness,
CPM signals are widely employed 
in aeronautical and space communication environments.
As an example, the IRIG 106 telemetry standard \cite{IRIG2004}
for control and non-payload communication (CNPC) links
employs several versions of CPM signals,
such as the PCM/FM, SOQPSK-TG, and ARTM formats.
Furthermore, the Radio Technical Commission for Aeronautics (RTCA)
is developing requirements and standards for CNPC links
\cite{RTCA2015}. To this aim, many transmission
techniques were evaluated for UAV 
communications: among them, one of the 
preferred solution is the GMSK format.
Moreover, adoption of GMSK has been proposed
also in deep space communications due to
the scarcity of spectrum for downlink transmission 
in the X-band \cite{Shambayati2012}.

\begin{figure}[t]
\centering
\includegraphics[width=0.90\columnwidth]{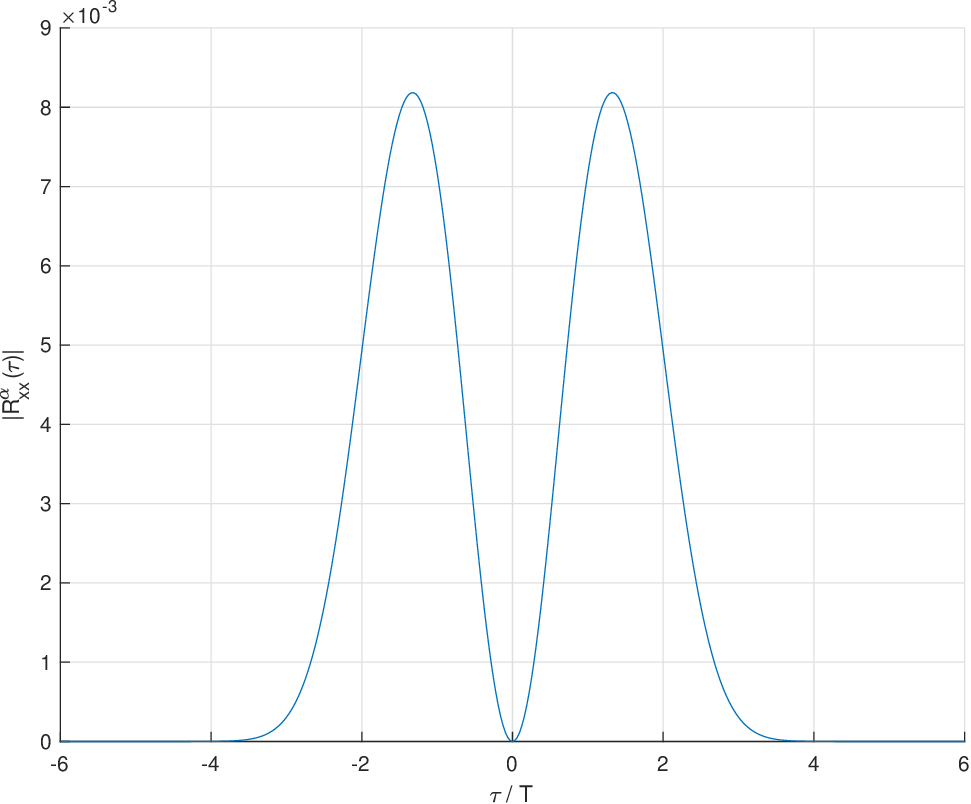}
\caption{\bf{Magnitude of the analytical 
cyclic ACF at $\alpha = 1/T$
for a GMSK signal.}}
\label{fig:cyclic-ACF-theo}
\end{figure}

The complex envelope
of a CPM signal having baud-rate $1/T$ 
may be expressed for $t \ge 0$ 
as \cite{Darsena2017}
\be
x(t) =
\exp\left[ j 2 \pi h
\sum_{n=0}^{+\infty} a_n \, g(t-n\, T)\right]
\label{eq:cpm}
\ee
where $h$ is the modulation index of the signal,
the information-bearing symbol sequence
$\{a_n\}_{n \ge 0}$ takes values in the $M$-ary alphabet
${\mathcal M}\eqdef\{ \pm 1,\pm 3,\ldots,\pm (M-1) \}$, $g(t)
\eqdef \int_{0}^{t} f(u)\, \text{d}u$ is the \textit{phase
response}, and $f(t)$ is the \textit{frequency response}
satisfying the following three conditions: 
\begin{enumerate}[(a)] \itemsep=2mm
    \item $f(t)\equiv0$ for each $t\not\in[0,LT]$;
    \item $f(t)=f(LT-t)$;
    \item $\int_{0}^{LT}f(u)\,\text{d}u=g(LT)=1/2$, with $L\in\Nset$.
\end{enumerate} 
The GMSK modulation format can be obtained as a particular case of
\eqref{eq:cpm} with $h = 1/2$ 
and a Gaussian-shaping pulse $f(t)$
with bandwidth $B$ \cite{Anderson2013}.

%===============================================================
%\section{Cyclic statistics of CPM signals}
%\label{sec:cpm_cyc}
%\subsection{Proper/improper}

%\textcolor{red}{[GG] Riporto qui qualche pezzo di testo da SOS-CPM con i riferimenti appropriati, che può %essere opportunamente modificato da Donatella}

The cyclic ACF and CCF of the CPM signal 
\eqref{eq:cpm} have been analytically evaluated in 
\cite{Darsena2017} by exploiting the 
Laurent representation.
It is shown in \cite{Darsena2017} that
the CPM signal is, in general, second-order 
wide-sense cyclostationary 
with cycle frequencies $\alpha = \frac{k}{T}$, 
with $k \in \Zset$.
In Fig.~\ref{fig:cyclic-ACF-theo}, we report the magnitude 
of the cyclic ACF \eqref{eq:cycACF-theo} 
evaluated at cycle frequency $\alpha = \frac{1}{T}$ for 
a \textit{binary} (i.e., $M=2$) GMSK signal having $h = 0.5$, $L=4$, and $B \, T = 0.25$. 

\begin{figure}[t]
\centering
\vspace{2mm}
\includegraphics[width=0.92\columnwidth]{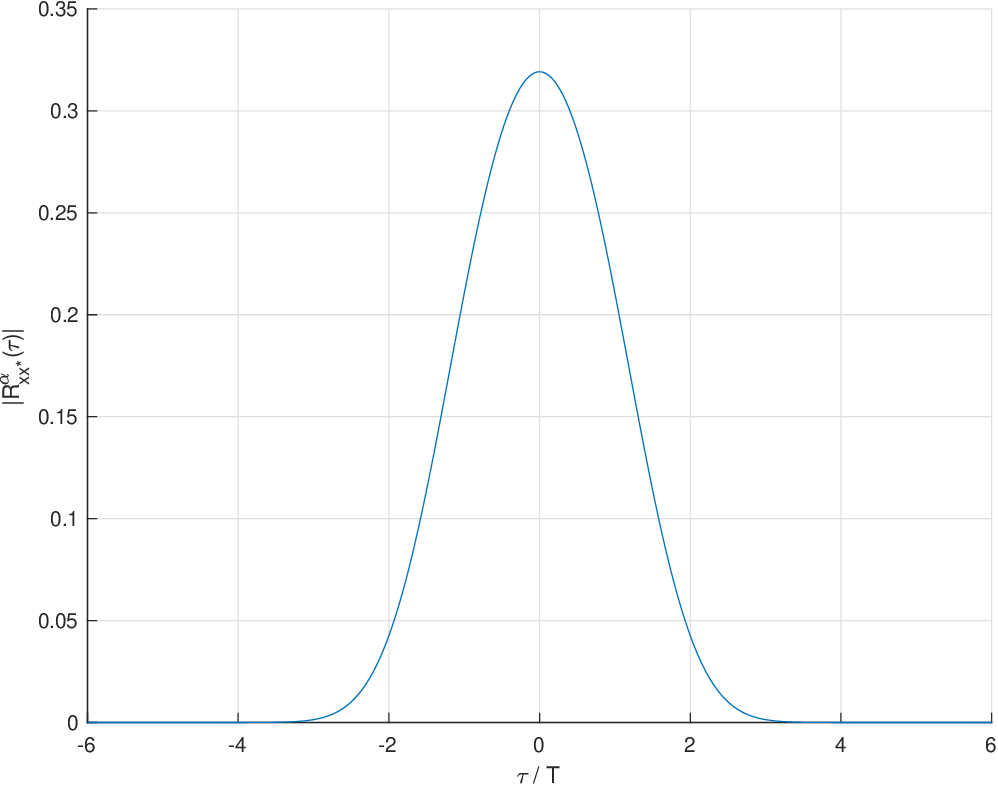}
\caption{\bf{Magnitude of the analytical 
cyclic CCF at $\beta = 1/2T$ 
for a GMSK signal.}}
\label{fig:cycCCF-theo}
\end{figure}

With reference to the cyclic conjugate statistics, instead, 
the statistical behavior of the CPM signal critically 
depends on the modulation index value. 
It can be proven, indeed, that when $h \neq \frac{1}{2}+k$, 
with $k \in \Zset$, the CCF turns out to be zero and, thus, the CPM signal 
is asymptotically \emph{circular} or \emph{proper} 
(see \cite{Neeser1993,Picinbono1994}).
On the contrary, for $h$ obeying $h = \frac{1}{2}+k$, with $k \in \Zset$,
the CPM signal is second-order wide-sense conjugate 
cyclostationary \cite{Darsena2017}, 
with conjugate cycle frequencies $\beta = \frac{1}{2T} + \frac{k}{T}$, 
as shown in Fig.~\ref{fig:cycCCF-theo}, 
where we report the magnitude of the cyclic CCF  \label{eq:cycCCF-theo}
at cycle frequency $\beta = \frac{1}{2T}$ for 
a binary GMSK signal having $h = 0.5$, $L=4$, and $B \, T = 0.25$.
%
%In this case, the CPM signal shows asymptotically non-vanishing
%\emph{noncircular}  or \emph{improper}
%\cite{Neeser1993,Picinbono1994} features.

%===============================================================
\section{GNU Radio implementation}
\label{sec:GR-implementation}

The SDR implementation has been carried 
out in the GNU Radio framework in order to simplify
further integration and, eventually, 
hardware deployment.
Specifically, the OOT module \code{CycloDSP}
was developed%
\footnote{The code is available at the GitHub repository
\url{https://github.com/krono-i2/gr-CycloDSP}}
, which contains 
the block \code{CycXCorr} implementing discrete-time 
estimation of the (conjugate or not) cyclic cross-correlation function
between two input signals.
In particular, when the same signal is applied to both inputs,
the block evaluates a discrete-time estimate 
of the cyclic ACF or CCF as defined in 
Section \ref{sec:ACS-signals}.

Since the \code{CycXCorr} block operates on continuous input streams,
in order to overcome the border effects 
introduced by windowing,
the cross-correlation function at time $n_0$ is evaluated as
\begin{multline}
\widehat{R}_{xy}^{\tilde{\alpha}}[n_0,m] =
\frac{1}{N} \sum_{n=0}^{N-1} x\left[n_0+\Mneg+n+m\right] \\
y^{*}\left[n_0+\Mneg+n\right] \, e^{-j 2 \pi \tilde{\alpha} n}
\label{eq:impl-corr}
\end{multline}
where $x[n]$ and $y[n]$ are the 
discrete-time input signals,
$m$ is the actual lag between inputs,
$N$ represents the length of the observation window,
$\tilde{\alpha}$ is the normalized cycle frequency,
and $\Mneg \eqdef \max_{m<0} |m|$.
A similar expression holds for the
conjugate cross-correlation function 
$\widehat{R}_{xy^*}^{\tilde{\beta}}[n_0,m]$
at the normalized conjugate cycle frequency 
$\tilde{\beta}$.

%\textcolor{red}{[GG] Qui avendo tolto il codice una serie
%di osservazioni che sono fatte nel testo mi sembra perdano significato, Ivan puoi
%aggiustare qui sotto e dovunque sia richiesto?}

The \code{CycXCorr} block was developed in Python%
\footnote{In particular, the implementation of the \code{CycXCorr} block 
is available at the link
\url{https://github.com/krono-i2/gr-CycloDSP/blob/master/python/CycloDSP/CycXCorr.py}.}
as a GNU Radio \emph{synchronous} block, i.e.,
which consumes and produces an equal number 
of items per port.
The block is characterized by \textit{base} 
attributes and \textit{specific}
attributes depending on the operation mode.
The base attributes are:
\code{mode}, an integer parameter controlling the operating mode;
\code{win\_len}, corresponding to $N$ in \eqref{eq:impl-corr};
\code{state}, storing the complex variable used to compensate
for the movement of the observation window;
\code{conj}, consisting on a boolean parameter controlling the conjugate
operator \eqref{eq:impl-corr}, i.e., when \code{conj} is \emph{False}
the cyclic ACF is evaluated, on the other hand,
when \code{conj} is \emph{True} the cyclic CCF is evaluated.
%
%The base attributes are:
%\begin{itemize}
%\item \code{mode}: it is an integer parameter controlling the operating mode.
%\item \code{win\_len}: it corresponds to $N$ in \eqref{eq:impl-corr}.
%\item \code{state}: it stores the complex variable used to compensate
%for the movement of the observation window.
%\item \code{conj}: it is a boolean parameter controlling the conjugate
%operator \eqref{eq:impl-corr}, i.e., when \code{conj} is \emph{False}
%the cyclic ACF is evaluated, on the other hand,
%when \code{conj} is \emph{True} the cyclic CCF is evaluated.
%\end{itemize}

When \code{mode} is set to $0$, the block works in
\emph{Set} mode, so the following specific attributes
arise: \code{max\_lag} corresponding to $\Mneg$ in \eqref{eq:impl-corr};
\code{alpha}, representing the list of required cycle frequencies
($\tilde{\alpha}$ or $\tilde{\beta}$).
%
%so the following specific attributes are added:
%
%\begin{itemize}
%\item \code{max\_lag}: it corresponds to $\Mneg$ in \eqref{eq:impl-corr}.
%\item \code{alpha}: it is the list of (conjugate) cycle frequencies
%$\tilde{\alpha}$ or $\tilde{\beta}$ it is needed to observe.
%\end{itemize}
%
Specifically, the cyclic (conjugate)
cross-correlation function is evaluated
for discrete lags $m$ spanning the range between 
\code{-max\_lag} and \code{max\_lag}, for all of the normalized 
cycle frequencies $\tilde{\alpha}$ or $\tilde{\beta}$ 
contained in the list \code{alpha}.

On the other hand, when \code{mode} is set to $1$,
the block works in \emph{Full} mode,
so the attribute \code{lags} needs to be 
added to the base attributes,
representing the list of required time 
lags $m$.
%
%so the following specific attributes are added:
%
%\begin{itemize}
%\item \code{lags}: it is the list of lags $m$ it is needed to observe.
%\end{itemize}
%
Specifically, the cyclic (conjugate) cross-correlation function
is evaluated for normalized cycle frequencies
$\tilde{\alpha}$ or $\tilde{\beta}$ spanning the range 
between $-1/2$ and $1/2$, for all discrete lags $m$ 
contained in the list \code{lags}.

%Other block parameters are \code{win\_len}, which  
%is an integer representing
%the length $N$ of the observation window, and \code{conj},
%which is a boolean variable used for switching between
%estimation of the cyclic cross-correlation function or
%that of the cyclic conjugate cross-correlation function.

%The input/output signature shows that the block
The \code{CycXCorr} block
accepts as inputs two complex scalar streams
and outputs one complex vector stream.
The dimension of the output items strictly depends
on the operation mode of the block, i.e.,
when in Set mode the block outputs complex vectors
of length \code{len(alpha)*(2*max\_lag+1)},
whereas when in Full mode the block 
outputs complex vectors
of length \code{len(lags)*win\_len}.

Moreover, when in Set mode, the block exploits
the function \code{scipy.signal.correlate}
from the open source Python library \emph{SciPy},
iterating on the set of normalized cycle frequencies.
In this case, the block needs to store the last
$N+2\Mpos$ samples of the inputs, with
$\Mpos \eqdef \max_{m>0} |m| = \Mneg$
(i.e., \code{max\_lag}).
Instead, when in Full mode, the block exploits the
\emph{Fast Fourier Transform} (FFT) algorithm
provided by the function \code{numpy.fft.fft}
included in the open source Python library \emph{NumPy},
iterating on the set of lags.
In this case, the block needs to store the last
$N+\max\left(\Mpos,\Mneg\right)$ samples of the inputs.
%
%For further details on the block implementation,
%please, refer to the source code on the GitHub repository%
%
%
It should be noted that the input buffer size
corresponds to the delay experienced by the output
with respect to the input streams.

Although we chose to implement 
as a first version
our algorithms entirely in Python,
in forthcoming releases of the software
we will perform code optimizations,
e.g., by resorting to C++ coding of the algorithm
and/or smarter implementations.
Finally, before implementing our algorithm on USRP hardware
(see Section \ref{sec:testbed}), we performed careful
loopback tests aimed at tuning the algorithms 
and properly setting their parameters.

%Finally, before implementing our algorithm with USRP hardware
%(see following Section), we carefully performed some
%loopback tests in order to fine tune the algorithms 
%and appropriately set their parameters.

%===============================================================
\section{Experimental results}
\label{sec:testbed}

In our experiments, we use two NI USRP-2920 devices
connected (with GbE cables) to two host computers 
both running GNU/Linux Ubuntu 22.04 and GNU Radio 3.10.1.1.
The first USRP generates a GMSK 
radio signal and transmits it over the air, 
whereas the second USRP, spaced approximately two meters apart, 
receives the GMSK radio signal, 
performs downconversion 
and sampling/quantization, and finally 
sends the I/Q samples to the host computer 
for cyclostationary signal analysis
through the \code{CycloDSP} module implemented in 
GNU Radio. All the experiments were carried out in  
a standard office room, 
exhibiting non-ideal propagation conditions.

\begin{figure*}[t]
\centering
\includegraphics[scale=0.12]{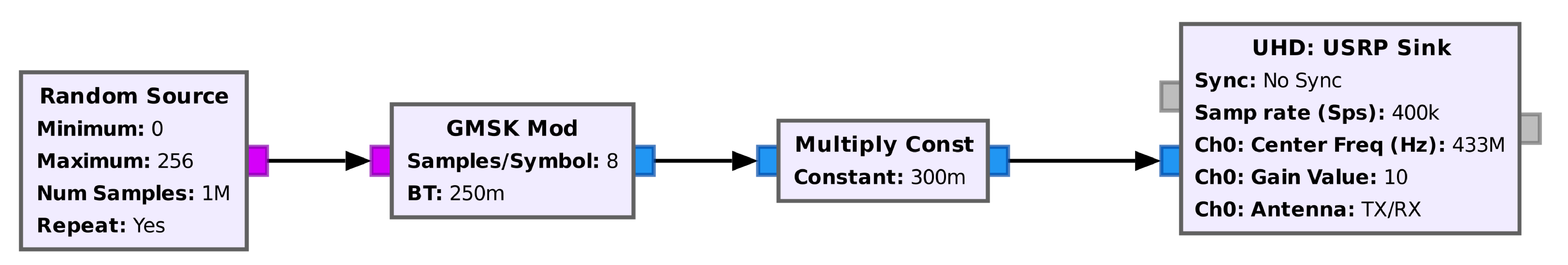}
\caption{\bf{GNU Radio Companion flowgraph of the GMSK transmitter.}}
\label{fig:grc-TX}
\end{figure*}

\begin{figure}[t]
\centering
\includegraphics[width=0.95\columnwidth]{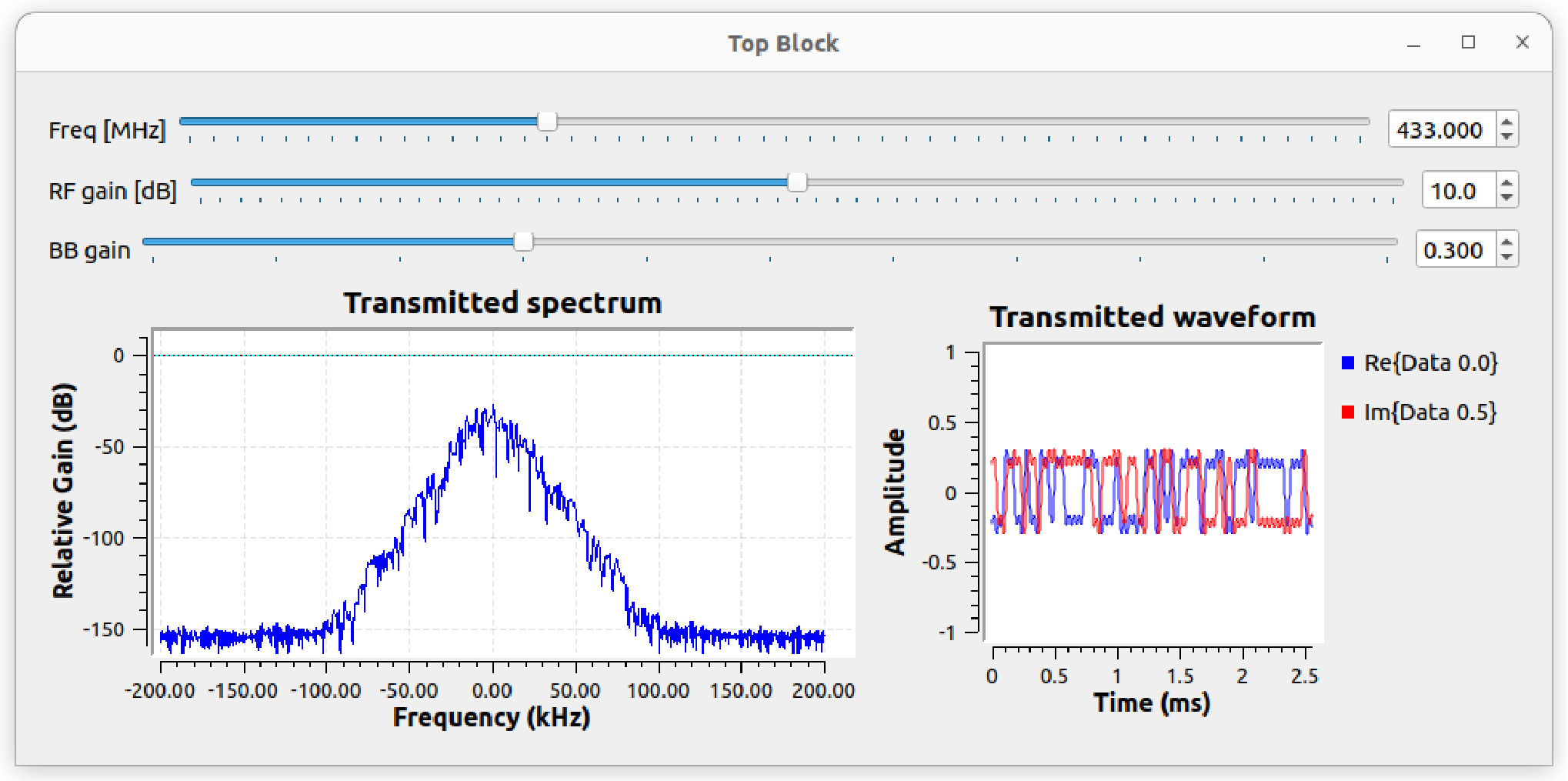}
\caption{\bf{GUI of the GMSK transmitter.}}
\label{fig:GUI-TX}
\end{figure}

At the transmitter side, we use standard GNU Radio blocks
in order to generate and send over the air through 
the USRP a binary GMSK signal.
A simplified version of the transmitter flowgraph 
is reported in Fig.~\ref{fig:grc-TX},
while Fig.~\ref{fig:GUI-TX} shows the \emph{graphical user interface}
(GUI) of the transmitter, where it is possible to set
the carrier frequency, the RF gain, and the baseband gain
of the transmitted signal; furthermore, by means of QT GUI 
blocks not reported in the flowgraph, 
it is possible to visualize the transmitted signal 
in both frequency- and time-domains.

\begin{figure*}[t]
\centering
\includegraphics[scale=0.12]{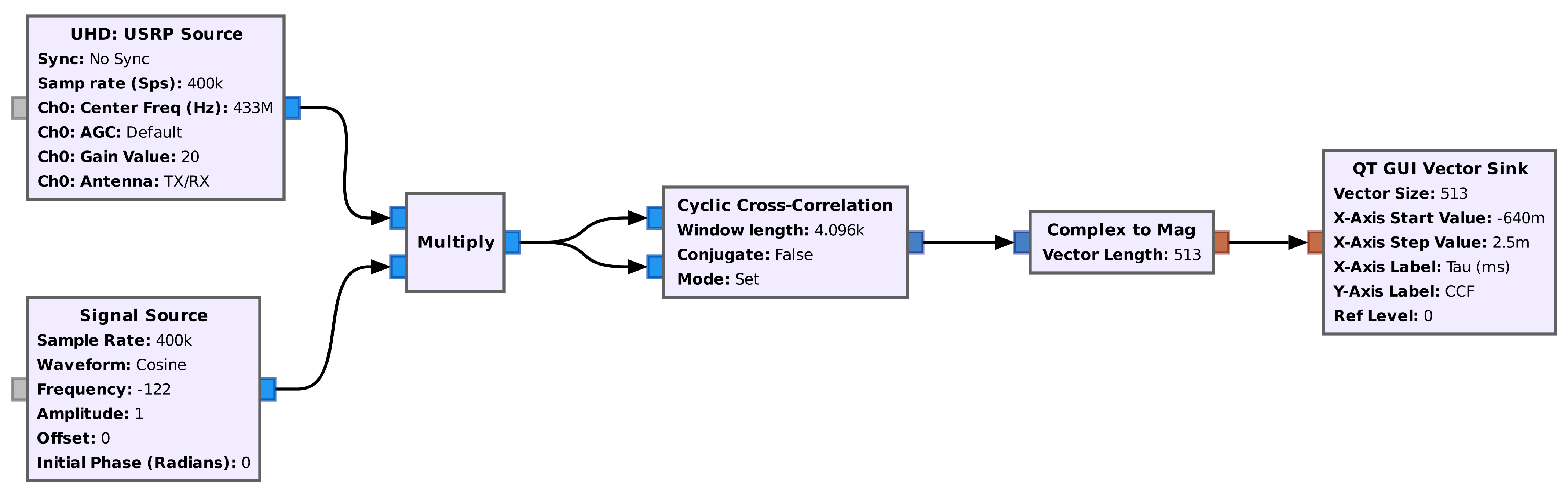}
\caption{\bf{GNU Radio Companion flowgraph of the receiver 
implementing cyclic ACF and CCF estimation.}}
\label{fig:grc-RX}
\end{figure*}

At the receiver, the host computer implements the flowgraph of
Fig.~\ref{fig:grc-RX}, aimed at performing the estimation 
of the cyclic ACF or cyclic CCF.
It should be noted that, before carrying out the experiments,
the two USRPs were carefully calibrated in order to compensate
for the {carrier frequency offset} (CFO) deriving from
the unavoidable differences of the oscillator clocks 
of the two RF frontends.
The CFO was preliminarily estimated and some blocks were included
in the receiving flowgraph in order to compensate for it
and allow correct estimation of the normalized cycle frequencies.
It is worthwhile to note that, although the CFO does not affect 
the cyclic ACF estimate, since it is automatically 
compensated by the conjugate operator in correlation evaluation,
such a compensation is needed when estimating 
the cyclic CCF,
since the CFO induces an offset on the 
normalized conjugate cycle frequency, which
would make the estimation completely 
unreliable.

\begin{table}[t]
    \centering
    \caption{\textbf{Main parameters of the GNU Radio experiments.}}
    \begin{tabular}{ll} 
     \toprule
     \textbf{Parameter} & \textbf{Value} \\
     \midrule
     Carrier frequency  & $433$ MHz \\
     Baud-rate          & $50$ kbaud \\
     $B \, T$           & 0.25 \\
     \# of samples/symbol & 8 \\
     Sample rate        & $400$ kHz \\ 
     \code{win\_len}     & $4096$ \\
     \code{mode}         & \emph{Set} \\
     \code{max\_lag}     & $256$ \\
    \bottomrule
    \end{tabular}
    \label{tab:simul-parameters}
\end{table}

\begin{figure}[t]
\centering
\includegraphics[width=0.95\columnwidth]{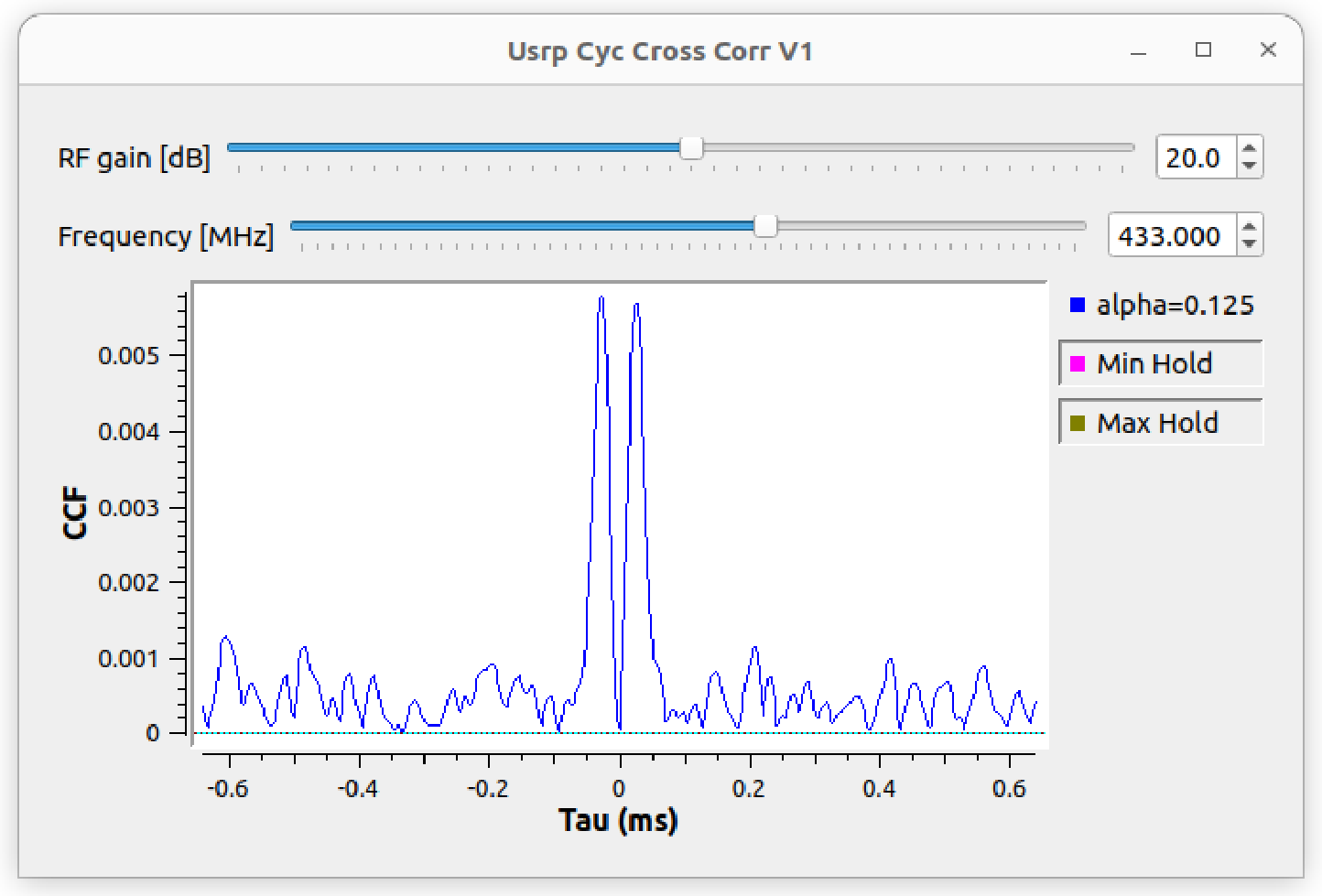}
\caption{\bf{GUI of the receiver for experiment \#1 (estimate of the cyclic ACF).}}
\label{fig:cyclic-ACF}
\end{figure}

Two experiments were carried out, whose
main parameters are reported in Tab.~\ref{tab:simul-parameters}.
In experiment \#1, the cyclic ACF is estimated
for $\alpha = 1/T$ and a corresponding normalized value
$\tilde{\alpha} = 0.125$.
In Fig.~\ref{fig:cyclic-ACF} we reported the GUI
of the receiver, showing a snapshot of the 
estimated cyclic ACF as a function
of the lag $\tau$. It is interesting to see that, 
unless side effects
due to the limited observation window and the non-ideal
radio enviroment, the estimated cyclic ACF is in close agreement 
with its theoretical counterpart
depicted in Fig.~\ref{fig:cyclic-ACF-theo}.

\begin{figure}[t]
\centering
\includegraphics[width=0.95\columnwidth]{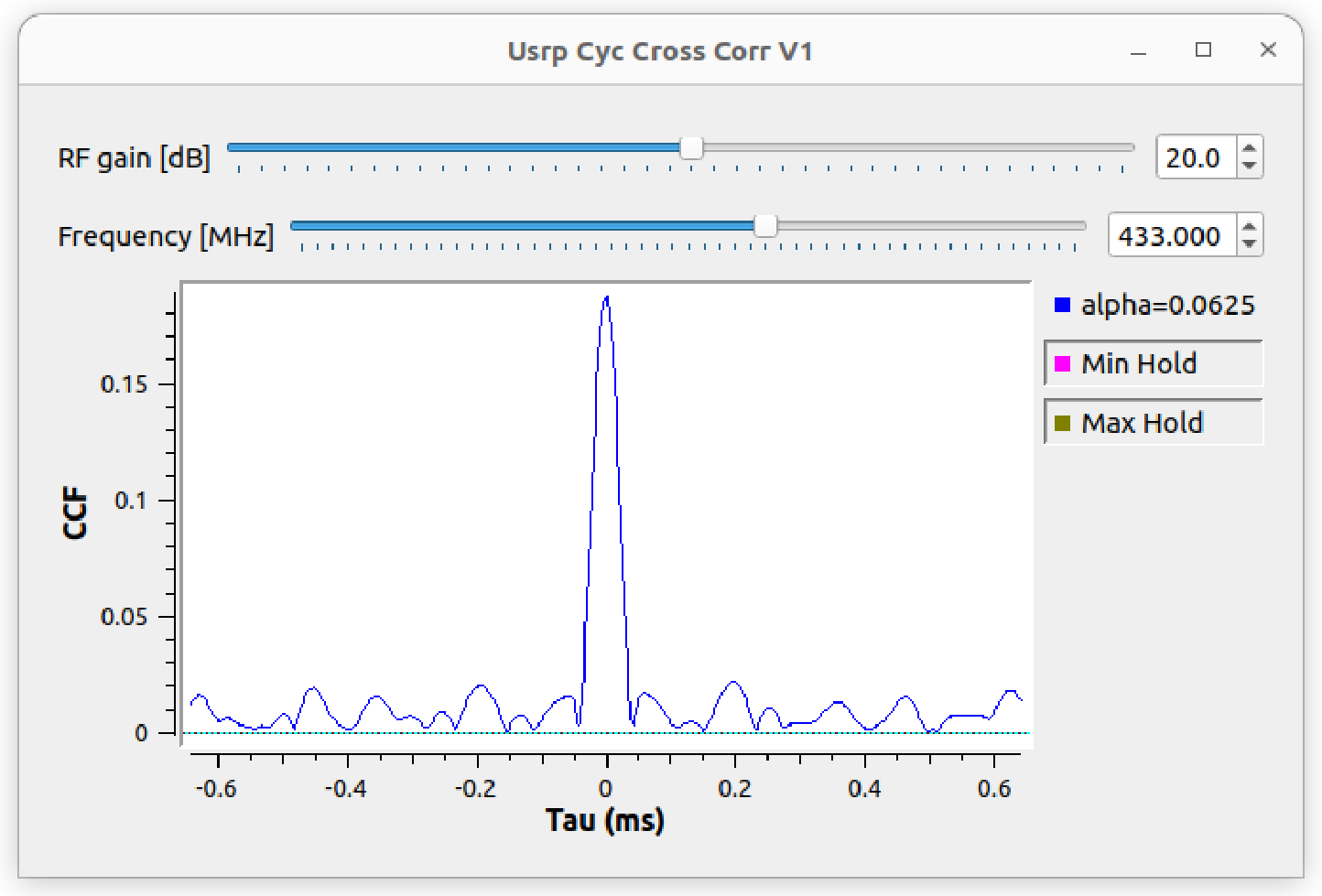}
\caption{\bf{GUI of the receiver for experiment \#2 (estimate of the cyclic CCF).}}
\label{fig:cyclic-CCF}
\end{figure}

In experiment \#2, the cyclic CCF is estimated
for $\beta = 1/2T$ and a corresponding normalized value
$\tilde{\beta} = 0.0625$.
In Fig.~\ref{fig:cyclic-CCF} we reported 
a snapshot of the estimated cyclic CCF as a function
of the lag $\tau$, which show again a close agreement 
with its theoretical counterpart 
depicted in Fig.~\ref{fig:cycCCF-theo}.

Finally, it should be remarked that, 
although the present version of the software 
is able to run in real-time even on modest-performance CPUs, 
it is not completely optimized, hence some functionalities 
of GNU Radio could not be completely available, e.g.,
interacting with the receiving USRP by means of the UHD driver
for changing parameters could suffer annoying lags.
Further optimization should be carried out
in order to reduce the computational burden
of the estimation algorithms.

%===============================================================
\section{Conclusion and future work}
\label{sec:concl}

%\textcolor{red}{[II] To be revised.}

In this paper, a first implementation of 
the OOT GNU Radio module \code{CycloDSP} for cyclostationary signal analysis
is described.
%This work represents the first step towards developing
%an extensible library of highly optimized algorithms
%for cyclostationary signal analysis, 
%both in time and frequency domains.
In particular, the module will contain a
set of building block functions 
to be used for real-time estimation
and visualization of the cyclic features 
of the input signal.
Although written in Python,
the algorithm is designed 
to work on continuous
data streams, and is able to be executed 
in real-time environments, even on modest-performance CPUs.
The \code{CycloDSP} module was validated by estimating 
in real-time the second-order cyclic statistics of a 
GMSK radio signal, widely utilized in aeronautical
and space applications. The test results are consistent with
the behavior expected by theory.

In forthcoming releases of the module, 
code optimization
will be performed by using specific 
hardware-oriented C++ libraries.
Furthermore, new functionalities will be implemented,
e.g., estimation of cyclic spectra and
cycle frequencies.

%===============================================================
\bibliographystyle{IEEEtran}
\bibliography{IEEEabrv,Gnuradio_v4}

\thebiography
%% This biostyle allows you to insert your photo size 1in X 1.25in

\begin{biographywithpic}
{Donatella Darsena}{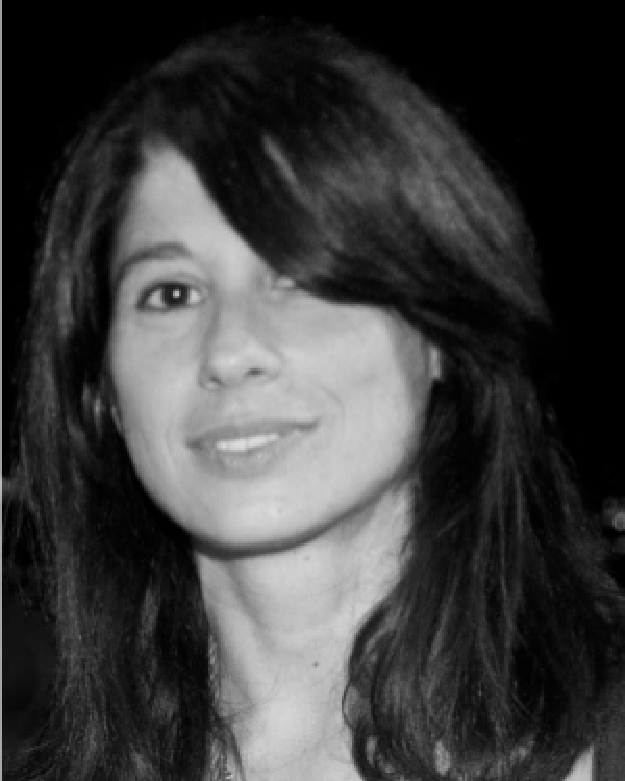} %(M'06-SM'16) 
received the Dr. Eng. degree
(\textit{summa cum laude}) in telecommunications engineering in 2001, and the Ph.D.
degree in electronic and telecommunications engineering in 2005,
both from the University of Napoli Federico II, Italy.
From 2001 to 2002, she was an engineer in the Telecommunications, Peripherals and
Automotive Group, STMicroelectronics, Milano, Italy.
From 2005 to 2022, she has been with the
Department of Engineering, University of Napoli Parthenope, Italy, where she served first 
as an Assistant Professor and then as an Associate Professor.
Since 2022, she has been an Associate Professor with the Department 
of Electrical Engineering and Information Technology, 
University of Napoli Federico II, Italy.
Her research activities lie in the area of statistical signal
processing, digital communications, and communication systems.
In particular, her current interests are focused on  
reflected-power communications, orthogonal/nonorthogonal multiple-access techniques, 
wireless systems optimization, and physical-layer security.

Prof. Darsena has served as Associate Editor for the IEEE SIGNAL PROCESSING LETTERS since 2020 
and Executive Editor for the IEEE COMMUNICATIONS LETTERS since 2023. 
She was an Associate Editor (from 2016 to 2019), Senior Area Editor (from 2019 to 2022), 
and Area Editor (from 2022 to 2023) of the IEEE COMMUNICATIONS LETTERS.
\end{biographywithpic}

\begin{biographywithpic}
{\noindent Giacinto Gelli}{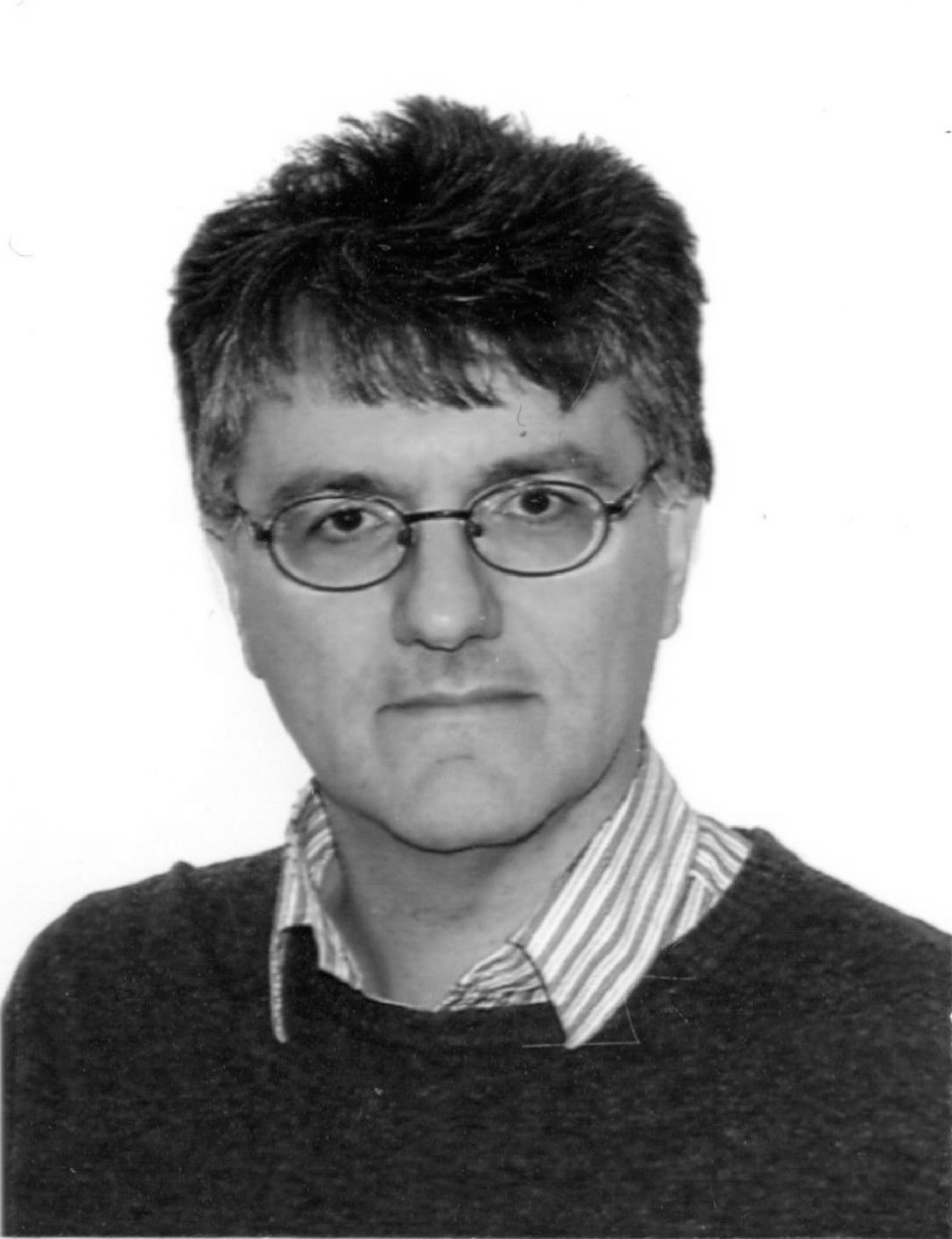} was born in Napoli, Italy, on July 29, 1964. 
He received the Dr. Eng. degree {\em summa cum laude} 
in electronic engineering in 1990, 
and the Ph.D. degree in computer science and
electronic engineering in 1994, 
both from the University of Napoli
Federico II, Italy.
From 1994 to 1998, he was an Assistant Professor with the
Department of Information Engineering, Second University of
Napoli, Italy. Since 1998 he has been with the Department 
of Electrical Engineering and Information Technology, 
University of Napoli Federico II, Italy, first as an Associate Professor,
and since November 2006 as a Full Professor of Telecommunications.
He also held teaching positions at the University 
Parthenope of Napoli, Italy.
His research interests are in the broad area of
signal and array processing for communications,
with current emphasis on multicarrier modulation systems,  
space-time techniques for cooperative and cognitive 
communications systems, and backscatter communications.
\end{biographywithpic} 

\begin{biographywithpic}
{Ivan Iudice}{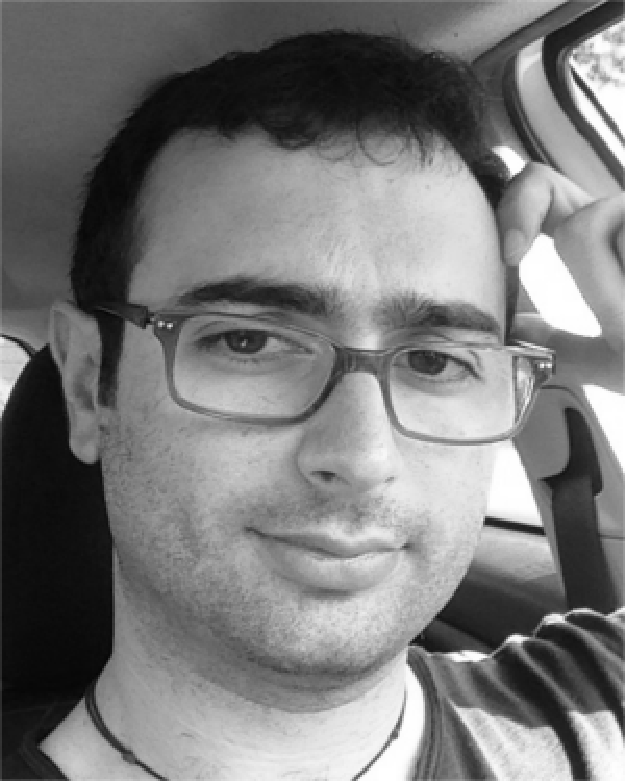}
was born in Livorno, Italy, on November 23, 1986.
He received the B.S. and M.S. degrees
in telecommunications engineering in 2008 and 2010, respectively,
and the
Ph.D. degree in information technology and electrical engineering in 2017,
all from University of Napoli Federico II, Italy.
Since 2011, he has been with
the Italian Aerospace Research Centre (CIRA), Capua, Italy.
He first served as part of the Electronics and Communications Laboratory and,
since November 2020, he has served as part of
the Security of Systems and Infrastructures Laboratory.
His research activities lie
in the area of signal and array processing for communications,
with current interests focused on physical-layer security,
space–time techniques for cooperative communications systems,
and software-defined radio.
\end{biographywithpic}

\end{document}